\documentclass{PoS}

\usepackage{amsmath}
\usepackage{subcaption}

\title{$\bar{b}\bar{b}ud$ tetraquark resonances in the Born-Oppenheimer approximation using lattice QCD potentials}

\ShortTitle{$\bar{b}\bar{b}ud$ tetraquark resonances}

\author{\speaker{Martin Pflaumer}$^a$, Pedro Bicudo$^b$, Marco Cardoso$^b$, Antje Peters$^a$, Marc Wagner$^a$ \\ \\
	$^a$Goethe-Universit\"at Frankfurt am Main, Institut f\"ur Theoretische Physik, \\ \phantom{xxxxx} Max-von-Laue-Stra{\ss}e 1, D-60438 Frankfurt am Main, Germany \\
	$^b$CeFEMA, Instituto Superior T\'{e}cnico, Universidade de Lisboa, Av.\ Rovisco Pais, \\ \phantom{xxxxx} 1049-001 Lisboa, Portugal \\ \\
		E-mail: \email{pflaumer@th.physik.uni-frankfurt.de}, \email{bicudo@tecnico.ulisboa.pt}, \email{mjdcc@cftp.ist.utl.pt}, \email{peters@th.physik.uni-frankfurt.de}, \email{mwagner@th.physik.uni-frankfurt.de}}

\abstract{We study tetraquark resonances using lattice QCD potentials for a pair of static antiquarks $\bar{b}\bar{b}$ in the presence of two light quarks $ud$. The system is treated in the Born-Oppenheimer approximation and we use the emergent wave method. We focus on the isospin $I=0$ channel, but consider different orbital angular momenta $l$ of the heavy antiquarks $\bar{b}\bar{b}$. We extract the phase shifts and search for $\mbox{S}$ and $\mbox{T}$ matrix poles on the second Riemann sheet. For orbital angular momentum $l=1$ we find a tetraquark resonance with quantum numbers $I(J^P)=0(1^-)$, resonance mass $m=10576^{+4}_{-4} \, \textrm{MeV}$ and decay width $\Gamma= 112^{+90}_{-103} \textrm{MeV}$, which can decay into two $B$ mesons.}

\FullConference{XIII Quark Confinement and the Hadron Spectrum - Confinement2018\\
		31 July - 6 August 2018\\
		Maynooth University, Ireland}

\newcommand{\rml}{\textit{\textrm{l}}}

\begin{document}


\section{Introduction}

A challenging and modern problem in particle physics and QCD is to improve our understanding of exotic hadrons. A possible approach to study heavy-heavy-light-light four-quark systems and the existence of tetraquarks is to compute potentials of two static antiquarks $\bar{Q}\bar{Q}$ in the presence of two light quarks $qq$ and to use these potentials in the Schr\"odinger equation to search for bound states (cf.\ e.g.\ \cite{Detmold:2007wk,Wagner:2010ad,Bali:2010xa,
Wagner:2011ev,Bicudo:2012qt,Brown:2012tm,
Bicudo:2015vta,Bicudo:2015kna,Bicudo:2016ooe}). In this way a stable $\bar{b}\bar{b}ud$ tetraquark with quantum numbers $I(J^P) = 0(1^+)$ has been predicted \cite{Bicudo:2012qt,Brown:2012tm}. Recently it has been confirmed by lattice computations using $\bar{b}$ quarks of finite mass treated with Non Relativistic QCD \cite{Francis:2016hui,Junnarkar:2018twb}. In this work, we extend our investigation of the $\bar{b}\bar{b}ud$ four-quark system by exploring the existence of tetraquark resonances. To this end we use the emergent wave method from scattering theory \cite{Bicudo:2015bra}.

For a more detailed discussion of this work cf.\ \cite{Bicudo:2017szl}.


\section{\label{sec:latticePotentials}Lattice QCD potentials of two static antiquarks $\bar{Q}\bar{Q}$ in the presence of two light quarks $qq$}

\begin{figure}[htb]
	\centering
	\includegraphics[width=0.45\textwidth]{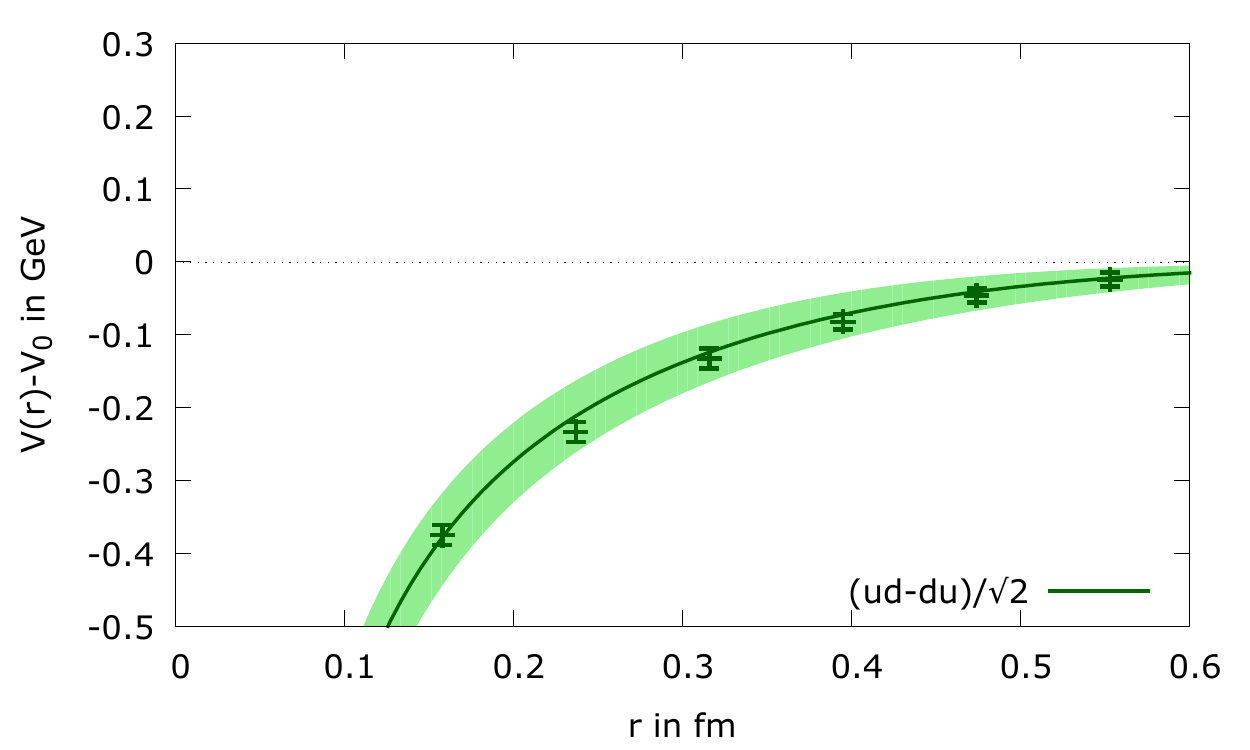}
	\includegraphics[width=0.45\textwidth]{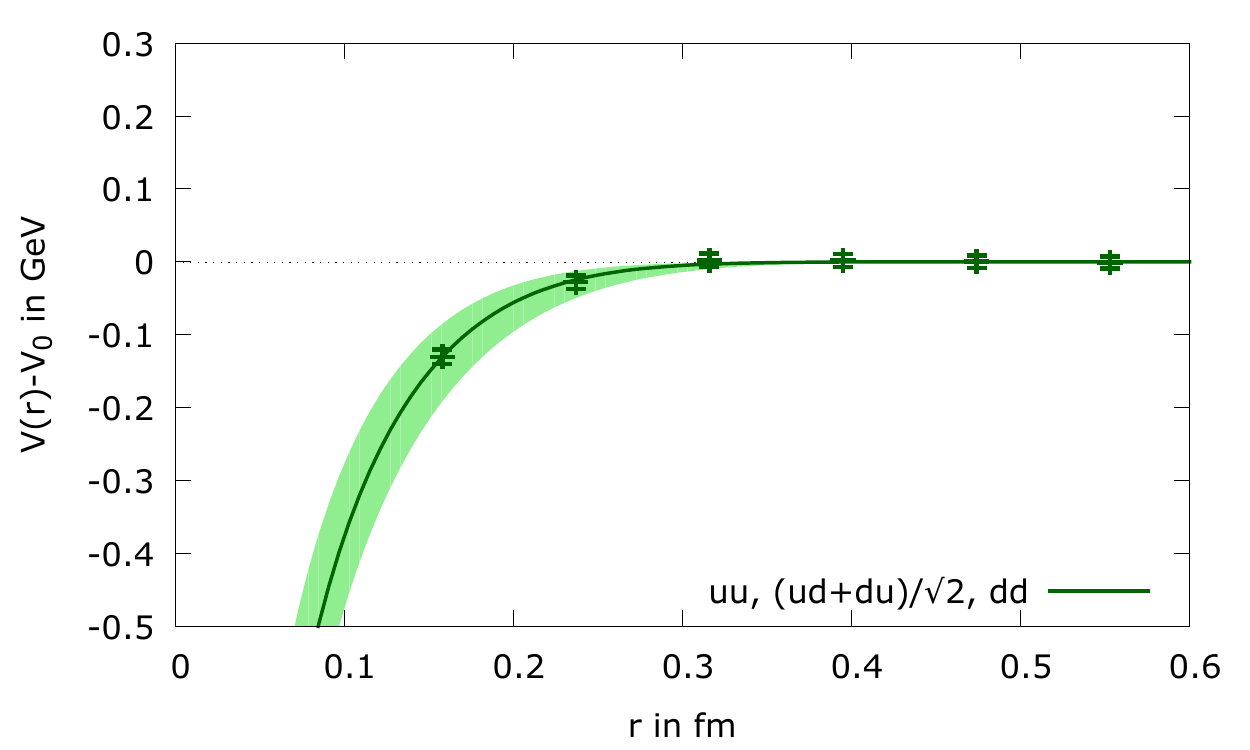}
	\caption{\label{fig:potentials}(left)~$(I=0,j=0)$ potential. (right)~$(I=1,j=1)$ potential.}
\end{figure}

In previous studies we have computed potentials $V(r)$ of two static antiquarks $\bar{Q} \bar{Q}$ in the presence of two light quarks $q q$ with lattice QCD. Computations have been performed for different light quark flavor combinations $q q$ with $q \in \{ u, d, s, c \}$. Moreover, different parity and total angular momentum sectors have been studied (cf.\ e.g.\ \cite{Bicudo:2015vta,Bicudo:2015kna}). There are both attractive as well as repulsive potentials. Of particular interest with respect to the existence of tetraquarks are two of the attractive potentials with $q \in \{ u , d \}$. The corresponding quantum numbers are $(I=0,j=0)$ and $(I=1,j=1)$, where $I$ denotes isospin and $j$ the total angular momentum of the light quarks and gluons around the $\bar{b}\bar{b}$ separation axis. The two potentials are shown in Figure~\ref{fig:potentials} for lattice spacing $a \approx 0.079 \, \textrm{fm}$ and $u$ and $d$ quark masses corresponding to a pion mass $m_\pi \approx 340 \, \textrm{MeV}$.

\begin{figure}[htb]
\centering
\includegraphics[width=0.75\columnwidth]{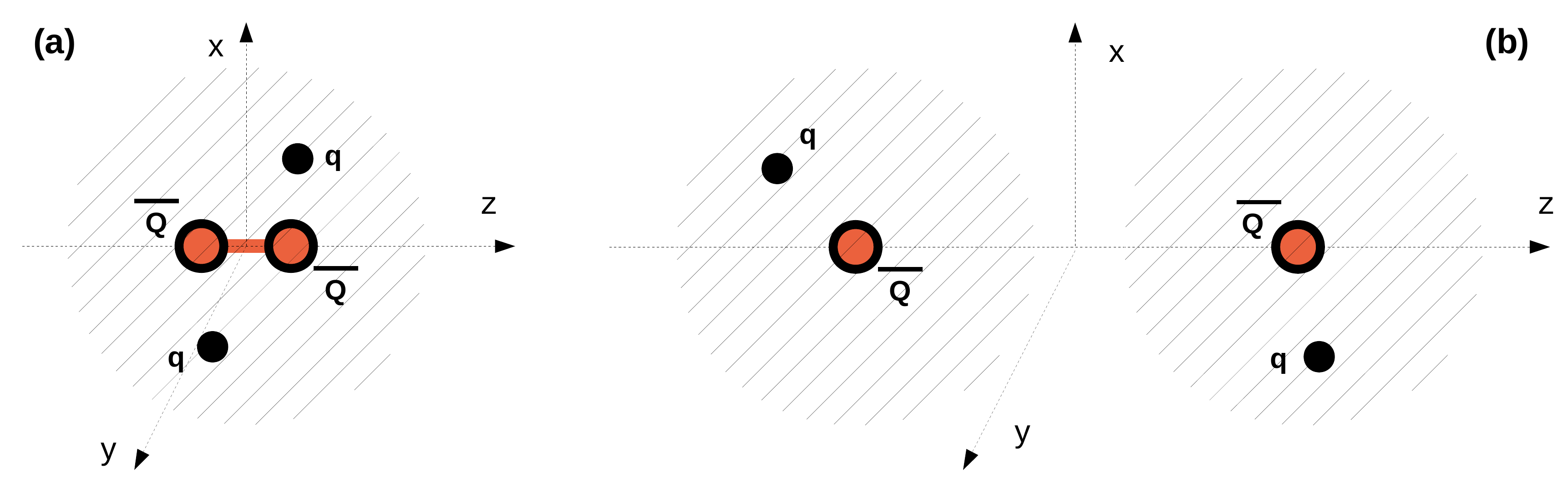}
\caption{\label{fig:screening}(a)~At small separations the static antiquarks $\bar{Q} \bar{Q}$ interact by perturbative one-gluon exchange. (b)~At large separations the light quarks $q q$ screen the interaction and the four quarks form two rather weakly interacting $B$ mesons.}
\end{figure}

The existence or non-existence of a stable tetraquark and its binding energy depends on the light quark mass $q$ \cite{Bicudo:2015vta}. Thus, we have performed computations of $\bar{Q} \bar{Q}$ potentials for three different light $u$ and $d$ quark masses corresponding to $m_\pi \in \{ 340 \, \textrm{MeV} , 480 \, \textrm{MeV} , 650 \, \textrm{MeV} \} $. The results, which can be parameterized by a screened Coulomb potential
\begin{equation}
\label{eq:potential} V(r) = -\frac{\alpha}{r} e^{-r^2 / d^2} ,
\end{equation}
have been extrapolated to $ m_\pi= 140 \, \textrm{MeV}$ \cite{Bicudo:2015kna}. The parameterization (\ref{eq:potential}) is motivated by one-gluon exchange for small $\bar{Q} \bar{Q}$ separations $r$ and the formation of two $B$ mesons at larger $r$ as a consequence of color screening as sketched in Figure~\ref{fig:screening}. Even though this approach is phenomenologically motivated, it is fully consistent with our lattice QCD results, i.e.\ the corresponding fits yield small $\chi^2 / \textrm{dof}$. The numerical values of the parameters $\alpha$ and $d$ are collected for both potentials in Table \ref{tab:parameters}. Clearly, the $(I = 0,j = 0)$ potential is more attractive than the $(I = 1,j = 1)$ potential.

\begin{table}[htb]
	\centering
	\begin{tabular}{cc|cc}
		\hline
		& & & \vspace{-0.4cm} \\
		$I$ & $j$ & $\alpha$ & $d$ in $\textrm{fm}$ \\
		& & & \vspace{-0.4cm} \\
		\hline
		& & & \vspace{-0.3cm} \\
		$\ 0 \ $ & $\ 0 \ $ & $\ 0.34^{+0.03}_{-0.03} \ $ & $\ 0.45^{+0.12}_{-0.10} \ $ \\
		& & & \vspace{-0.25cm} \\
		$1$ & $1$ & $0.29^{+0.05}_{-0.06}$ & $0.16^{+0.05}_{-0.02}$\vspace{-0.3cm} \\
		& & & \\
		\hline
	\end{tabular}
\caption{\label{tab:parameters}Parameters $\alpha$ and $d$ of the parameterization (\ref{eq:potential}) of the two attactive $\bar{Q} \bar{Q}$ potentials with $(I = 0,j = 0)$ and $(I = 1,j = 1)$.}
\end{table}

The potential parameterization (\ref{eq:potential}) with $\alpha$ and $d$ from Table~\ref{tab:parameters} can be inserted into the Schr\"o\-dinger equation, i.e.\ they can be used to explore the existence of stable tetraquarks or tetraquark resonances in the Born-Oppenheimer approximation. A stable $\bar{b} \bar{b} u d$ tetraquark with quantum numbers $I(J^P) = 0(1^+)$ around $60 \, \textrm{MeV}$ below the $B B^\ast$ threshold has been predicted in \cite{Bicudo:2012qt,Bicudo:2015vta,Bicudo:2015kna,Bicudo:2016ooe}. The search for $\bar{b} \bar{b} u d$ tetraquark resonances is discussed in \cite{Bicudo:2017szl} and the following sections.


\section{\label{sec:emergent_wave}The emergent wave method}

In this section we discuss the emergent wave method, which allows to extract phase shifts and resonance parameters. More details can be found e.g.\ in \cite{Bicudo:2015bra}.

We start by considering the Schr\"odinger equation
\begin{equation}
\label{eq:schro} \Big(H_{0} + V(r)\Big) \Psi = E \Psi
\end{equation}
and by splitting the wave function into two parts,
\begin{equation}
\label{eq:sep_psi} \Psi = \Psi_{0} + X .
\end{equation}
$\Psi_{0}$ is the incident wave, which is a solution of the free Schr\"odinger equation, i.e.\ $H_{0} \Psi_{0} = E \Psi_{0}$, and $X$ denotes the emergent wave. Inserting eq.\ (\ref{eq:sep_psi}) into eq.\ (\ref{eq:schro}) and using the free Schr\"odinger equation we obtain
\begin{equation}
\label{eq:schro_scatter} \Big(H_{0} + V(r) - E\Big) X = -V(r) \Psi_{0} .
\end{equation}
Solving this equation for given energy $E$ provides the emergent wave $X$. The asymptotic behavior of $X$ determines the phase shifts. We find the poles of the $\mbox{S}$ matrix and the $\mbox{T}$ matrix in the complex energy plane and identify them with resonances, when located in the second Riemann sheet at $\mathcal{E} - i \Gamma/2$, where $\mathcal{E}$ is the energy and $\Gamma$ is the decay width of the resonance.


\subsection{Partial wave decomposition}

The Hamiltonian describing the two heavy antiquarks $\bar{b} \bar{b}$ is
\begin{equation}
\label{EQN005} H = H_0 + V(r) = -\frac{\hbar^{2}}{2 \mu} \triangle + V(r) ,
\end{equation}
where $\mu = M/2$ is the reduced mass and $M = 5 \, 280 \, \textrm{MeV}$ is the mass of the $B$ meson from the PDG \cite{Tanabashi:2018oca}. One can express the incident plane wave $\Psi_{0} = e^{i \mathbf{k} \cdot \mathbf{r}}$ as a sum over spherical waves,
\begin{equation}
\label{eq:expansionsphericalbessel} \Psi_{0} = e^{i \mathbf{k} \cdot \mathbf{r}} = \sum_{l} (2l+1) i^{l} j_{l}(k r) P_{l}(\hat{\mathbf{k}} \cdot \hat{\mathbf{r}}) ,
\end{equation}
where $j_{l}$ are spherical Bessel functions, $P_{l}$ are Legendre polynomials and the relation between energy and momentum is $\hbar k = \sqrt{2 \mu E}$. Since the potential $V(r)$ is spherically symmetric, we can also expand the emergent wave $X$ in terms of Legendre polynomials $P_{l}$, 
\begin{equation}
\label{eq:001} X = \sum_{l} (2l+1) i^{l} \frac{\chi_l(r)}{k r} P_{l}(\hat{\mathbf{k}} \cdot \hat{\mathbf{r}}) .
\end{equation}
Inserting eq.\ (\ref{eq:expansionsphericalbessel}) and eq.\ (\ref{eq:001}) into eq.\ (\ref{eq:schro_scatter}) leads to a set of ordinary differential equations for $\chi_l$,
\begin{equation}
\label{eq:1cl0:radial} \bigg(-\frac{\hbar^2}{2 \mu} \frac{d^{2}}{dr^{2}} + \frac{l (l+1)}{2 \mu r^{2}} + V(r) - E\bigg) \chi_l(r) = -V(r) k r j_l(k r) .
\end{equation}


\subsection{Solving the differential equations for the emergent wave}

$V(r)$, eq.\ (\ref{eq:potential}), is exponentially screened, i.e.\ $V(r) \approx 0$ for $r \geq R$, where $R \gg d$. Consequently, the emergent wave is a superposition of outgoing spherical waves for large separations $r \geq R$ and can be expressed by spherical Hankel functions of the first kind $h_l^{(1)}$,
\begin{equation}
\label{eq:002} \frac{\chi_l(r)}{k r} = i t_l h_l^{(1)}(k r) .
\end{equation}
To compute the complex prefactors $t_l$, which will lead to the phase shifts, we solve the ordinary differential equation (\ref{eq:1cl0:radial}) using the following boundary conditions:
\begin{itemize}
\item At $r = 0$: $\chi_l(r) \propto r^{l+1}$.
	
\item For $r \geq R$: eq.\ (\ref{eq:002}).
\end{itemize}
We emphasize that the boundary condition for $r \geq R$ depends on $t_l$. Solving the differential equation for a given value of the energy $E$, this boundary condition is only fulfilled for a specific value of $t_l$. In other words the boundary condition for $r \geq R$ fixes $t_l$ as a function of $E$.

To solve eq.\ (\ref{eq:1cl0:radial}) numerically, we have implemented two different approaches:
\begin{itemize}
\item [(1)] A fine uniform discretization of the interval $[0,R]$ reducing the differential equation to a large set of linear equations, which can be solved rather efficiently, since the corresponding matrix is tridiagonal.

\item [(2)] A standard 4-th order Runge-Kutta shooting method.
\end{itemize}


\subsection{Phase shifts, $\mbox{S}$ and $\mbox{T}$ matrix poles}

$t_l$ is an eigenvalue of the $\mbox{T}$ matrix (see standard textbooks on quantum mechanics and scattering, e.g.\ \cite{Merzbacher}). From $t_l$ we can determine the phase shift $\delta_l$ and also the corresponding $\mbox{S}$ matrix eigenvalue
\begin{equation}
\label{eq:003} s_l \equiv 1 + 2 i t_l = e^{2 i \delta_l}
\end{equation}
(at large distances $r \geq R$ the radial wave function is $k r ( j_l (kr) + i t_l h_l^{(1)}(k r)) =	(k r /2) (h_l ^{(2)}(kr) + e^{2 i \delta_l} h_l^{(1)}(k r))$). Note that both the $\mbox{S}$ matrix and the $\mbox{T}$ matrix are analytical in the complex plane and are also defined for complex energies $E$. Thus, we solve the differential equation (\ref{eq:1cl0:radial}) for complex $E$ and find the $\mbox{S}$ and $\mbox{T}$ matrix poles by scanning the complex energy plane $(\textrm{Re}(E) , \textrm{Im}(E))$ and by applying Newton's method to find the roots of $1 / t_l(E)$. These poles correspond to complex resonance energies $E = \mathcal{E} - i \Gamma/2$ and must be located in the second Riemann sheet with a negative imaginary part of $E$.


\section{\label{sec:results}Results for phase shifts, $\mbox{S}$ and $\mbox{T}$ matrix poles and prediction of resonances}


\subsection{Phase shifts $\delta_l$}

\begin{figure}
	\centering
	\includegraphics[width=0.48\textwidth]{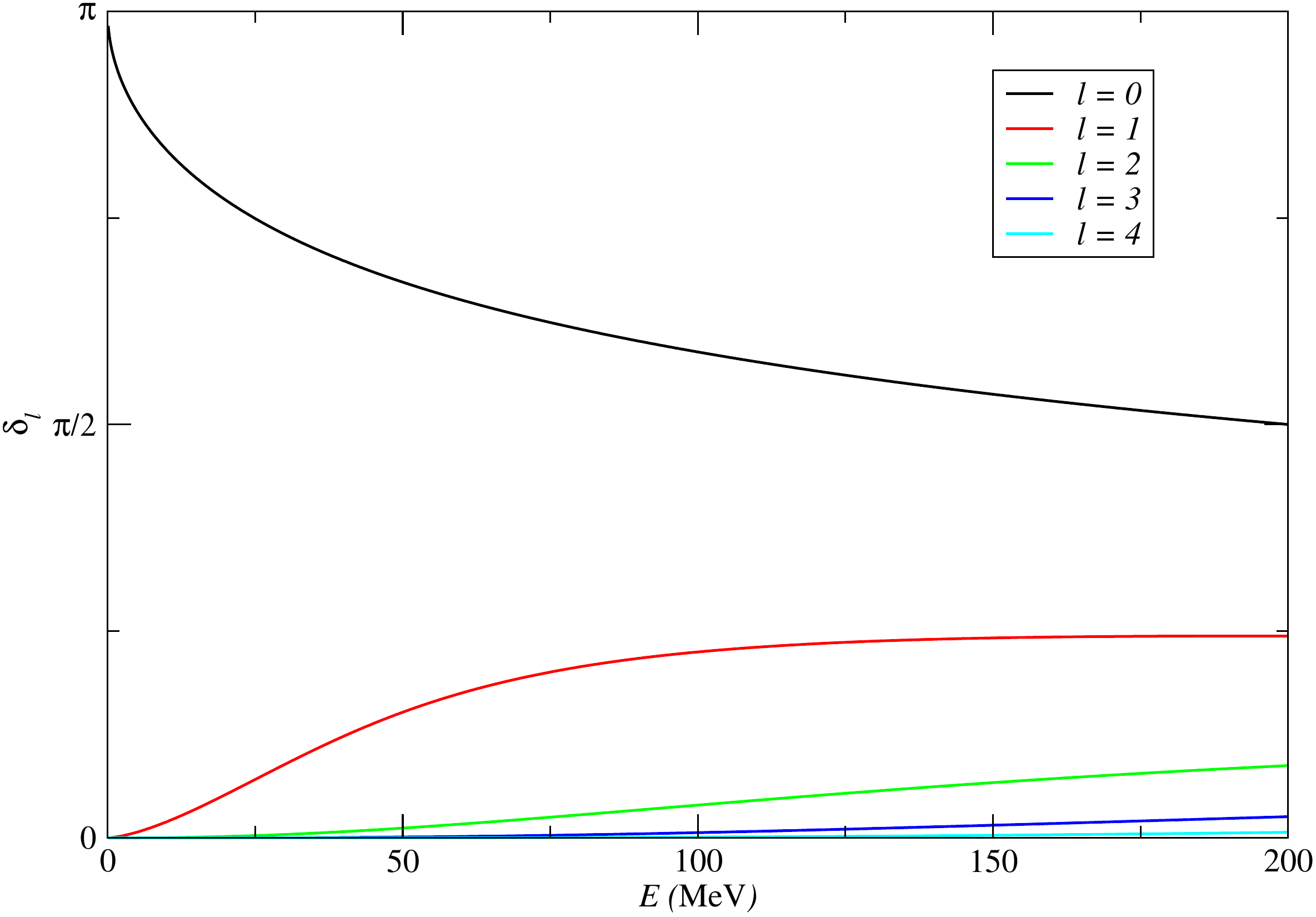}
	\includegraphics[width=0.48\textwidth]{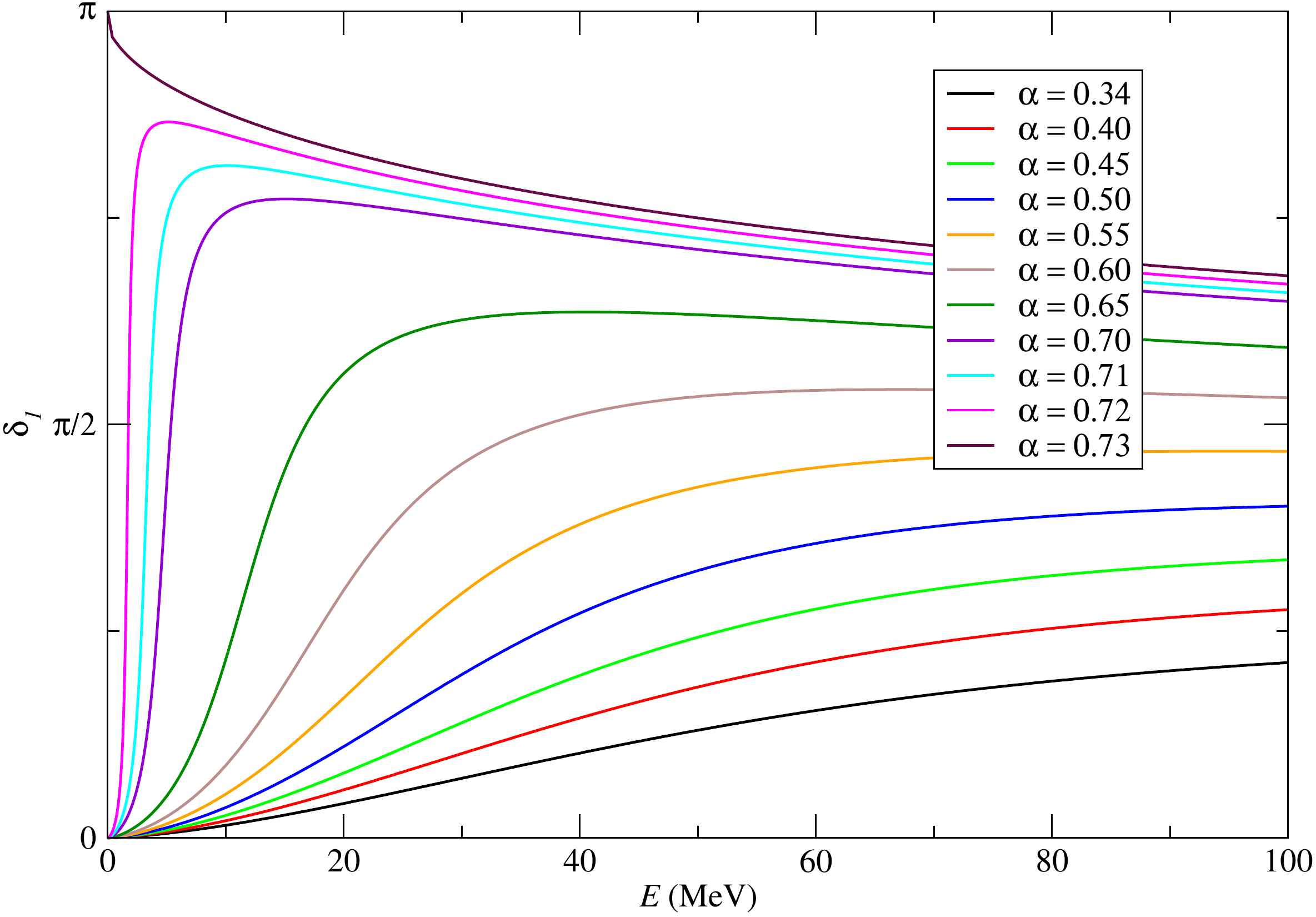}\\
	\caption{\label{fig:phase_shifts}(left)~Phase shift $\delta_l$ as a function of the energy $E$ for orbital angular momenta $l=0,1,2,3,4$ for the $(I=0,j=0)$ potential ($\alpha = 0.34$, $d = 0.45 \, \textrm{fm}$). (right)~Phase shift $\delta_1$ as a function of the energy $E$ for different $\alpha$ and fixed $d = 0.45 \, \textrm{fm}$.}
\end{figure}

We consider the more attractive $\bar{b} \bar{b}$ potential with $(I=0,j=0)$ (cf.\ section~\ref{sec:latticePotentials}), compute $t_l$ for real energies $E$ and apply eq.\ (\ref{eq:003}) to determine phase shifts $\delta_l$ for orbital angular momenta $l=0,1,2,3,4$. A clear indication for a resonance would be a strongly increasing $\delta_{\rml}$ from $0$ to almost $\pi$. Such a behavior is, however, not observed (cf.\ Figure~\ref{fig:phase_shifts} (left)). Thus, we have to check more thoroughly, whether there are resonances or not.

It is also interesting to consider the $l=1$ channel for even more attractive potentials by increasing the parameter $\alpha$, while $d$ is fixed. We show the resulting phase shifts $\delta_1$ in Figure~\ref{fig:phase_shifts} (right). For $\alpha \gtrsim 0.65$ resonances are clearly indicated. For $\alpha \gtrsim 0.72$ there are even bounds states, i.e.\ the phase shifts start at $\pi$ and decrease monotonically. However, this observation does not allow to make a clear statement, whether there is a resonance for $\alpha = 0.34$.


\subsection{Resonances as poles of the $\mbox{S}$ and $\mbox{T}$ matrices for complex energies $E$}

\begin{figure}
	\centering
	\includegraphics[trim=35 25 30 35, clip,width=0.48\textwidth]{complex_plane.pdf}
	\includegraphics[trim=35 25 10 0, clip, width=0.48\textwidth]{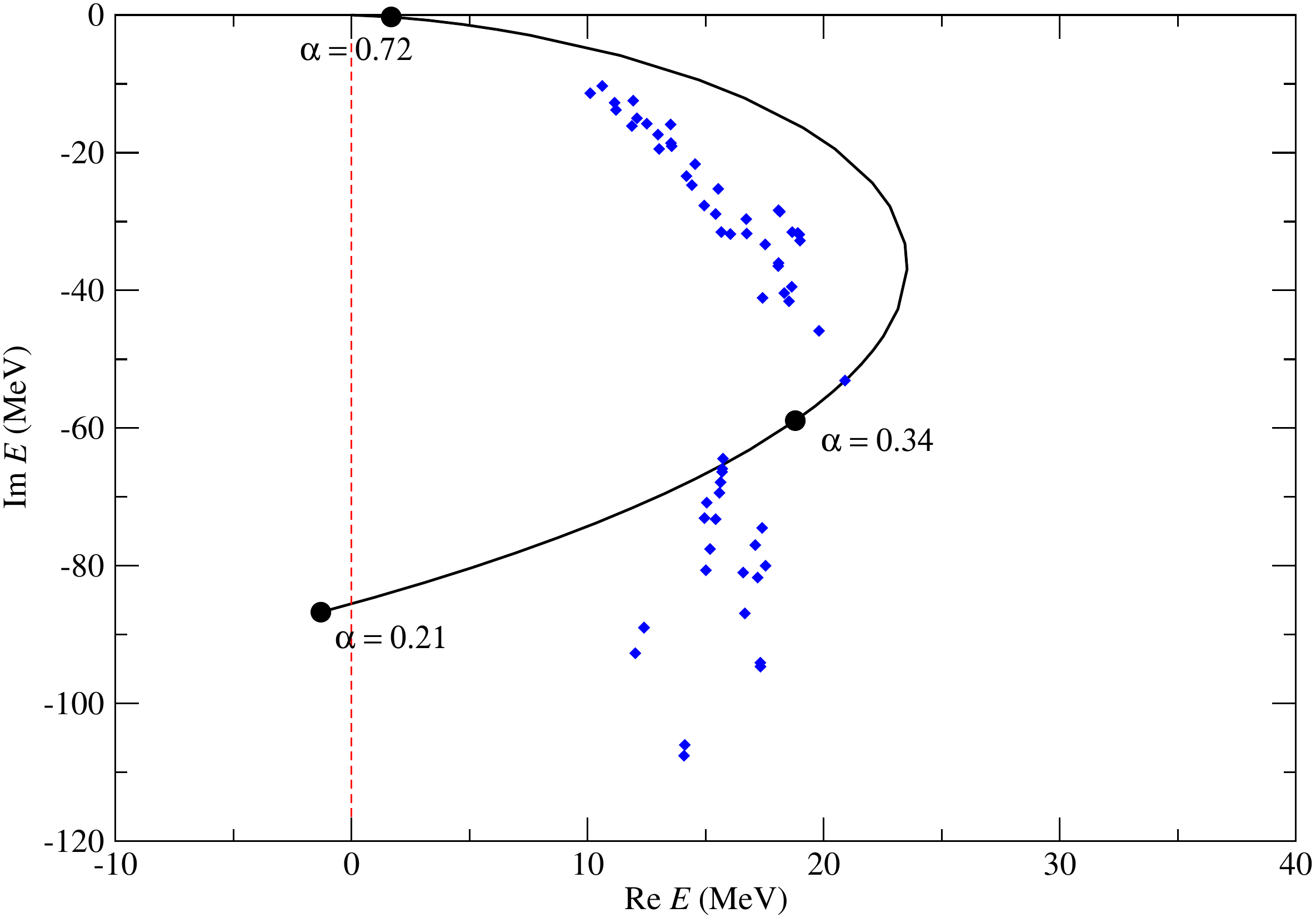}
	\caption{\label{fig:complexPlane}(left):~$\mbox{T}$ matrix eigenvalue $t_1$ as a function of the complex energy $E$ for the $(I=0,j=0)$ potential ($\alpha = 0.34$, $d = 0.45 \, \textrm{fm}$). Along the vertical axis we show the norm $|t_1|$, while the phase $\textrm{arg}(t_1)$ is visualized by different colors.
	(right)~Trajectory of the pole of the $\mbox{T}$ matrix eigenvalue $t_1$ in the complex energy plane $(\textrm{Re}(E),\textrm{Im}(E))$ corresponding to a variation of the parameter $\alpha$. The cloud of blue points represents the systematic error of our prediction.}
\end{figure}

Now we search for poles of the $\mbox{T}$ matrix eigenvalue $t_l$ in the complex energy plane, which indicate resonances. For orbital angular momentum $l=1$ and the $(I=0,j=0)$ potential we find a pole, which is shown in Figure~\ref{fig:complexPlane} (left), where $t_1$ is plotted as a function of the complex energy $E$. For a better understanding of the resonance and its dependence on the potential we determine the pole of $t_1$ for various parameters $\alpha$. In Figure~\ref{fig:complexPlane} (right) we show the location of the pole for several values of $\alpha$ in the  $(\textrm{Re}(E),\textrm{Im}(E))$ plane. Indeed, starting at $\alpha=0.21$ we find poles. Consequently, we can predict a resonance at $\alpha =0.34$. For orbital angular momenta $l > 1$ as well as for the less attractive potential $(I=1,j=1)$ no poles have been found.

\subsection{Analysis of statistical and systematic errors}

We perform a detailed statistical and systematic error analysis for the pole of $t_1$ in the complex energy plane $(\textrm{Re}(E),\textrm{Im}(E))$ using the same method as for our study of bound states \cite{Bicudo:2015vta}. We parametrize the lattice QCD data for the potential $V^{\textrm{lat}}(r)$ with an uncorrelated $\chi^2$ minimizing fit using the ansatz (\ref{eq:potential}), i.e.\ we minimize the expression
\begin{equation}
\label{eq:chisquared} \chi^2 = \sum_{r_\textrm{min} \leq r \leq r_\textrm{max}} \bigg(\frac{V(r)-V^{\textrm{lat}}(r)}{\Delta V^{\textrm{lat}}(r)}\bigg)^2
\end{equation}
with respect to $\alpha$ and $d$, where $\Delta V^{\textrm{lat}}(r)$ denotes the corresponding statistical errors. To estimate the systematic error, we perform fits for various fit ranges $r_\textrm{min} \leq r \leq r_\textrm{max}$. Additionally, we vary the range of the temporal separation $t_\textrm{min} \leq t \leq t_\textrm{max}$, where $V^{\textrm{lat}}(r)$ is read off. For each fit we determine the pole of $t_1$, i.e.\ the resonance energy $\mathcal{E}$ and the decay width $\Gamma$. As systematic error we take the spread of these results, while the statistical error is determined via the jackknife method. Applying this combined systematic and statistical error analysis, we find a resonance energy $\mathcal{E} = \textrm{Re}(E) = 17^{+4}_{-4} \, \textrm{MeV}$ above the $B B$ threshold and a decay width $\Gamma = -2 \textrm{Im}(E) = 112^{+90}_{-103} \, \textrm{MeV}$. Studying the symmetries of the quarks with respect to color, flavor, spin and their spatial wave function and considering the Pauli principle we determine the quantum numbers as $I(J^P) = 0(1^-)$. The mass of this $\bar{b}\bar{b}ud$ tetraquark resonance is given by $m = 2 M + \textrm{Re}(E) = 10 \, 576^{+4}_{-4} \, \textrm{MeV}$.


\section{Conclusion}

We have explored the existence of $\bar{b}\bar{b}ud$ tetraquark resonances applying lattice QCD potentials for two static antiquarks in the presence of two light quarks, the Born-Oppenheimer approximation and the emergent wave method. We predict a new resonance with quantum numbers $I(J^P)=0(1^-)$, a resonance mass $\textrm{Re}(E) = 17^{+4}_{-4} \, \textrm{MeV}$ and a decay width $\Gamma  = 112^{+90}_{-103} \, \textrm{MeV}$.


\section*{Acknowledgements}

We acknowledge useful conversations with K.~Cichy.

P.B.\ acknowledges the support of CeFEMA (grant FCT UID/CTM/04540/2013) and is thankful for hospitality at the Institute of Theoretical Physics of Goethe-University Frankfurt am Main. M.C.\ acknowledges the support of CeFEMA and the FCT contract SFRH/BPD/73140/2010. M.W.\ acknowledges support by the Emmy Noether Programme of the DFG (German Research Foundation), grant WA 3000/1-1.

This work was supported in part by the Helmholtz International Center for FAIR within the framework of the LOEWE program launched by the State of Hesse.

Calculations on the LOEWE-CSC and on the on the FUCHS-CSC high-performance computer of the Frankfurt University were conducted for this research. We would like to thank HPC-Hessen, funded by the State Ministry of Higher Education, Research and the Arts, for programming advice.



\end{document}